\documentstyle[preprint,aps]{revtex}
\def\IR{{\hbox{{\rm I}\kern-.2em\hbox{\rm R}}}}
\newcommand{\be}{\begin{equation}}
\newcommand{\ee}{\end{equation}}

\begin{document}
\preprint{\begin{tabular}{l}
           IASSNS-HEP-95/43 \\
           hep-th/9510180
           \end{tabular}
         }

\title{Statistical
Thermodynamics of a Black Hole\\ in Terms of Surface Fields \footnote {A
preliminary version of this paper has been cited under the title,
``Action principle, state counting and statistical thermodynamics of
black holes''. }}
\addtocounter{footnote}{1}

\author {Claudio Teitelboim \thanks{Electronic address: teitel@cecs.cl}}
\address{ Centro de Estudios Cient\'{\i}ficos de Santiago,
Casilla 16443, Santiago 9, Chile\\ and \\
Institute for Advanced Study, Olden Lane, Princeton, NJ 08540, USA}

\date{\today}
\maketitle

\begin{abstract}

An action principle for spacetimes with the topology of an Euclidean
black-hole is given.  The gravitational field is described by the
ordinary volume degrees of freedom plus additional surface fields at
the horizon.  The surface degrees of freedom correspond to
diffeomorphisms on the sphere at the horizon and a field of ``opening
angles''.  General covariance forces the surface modes to be confined
to a box of an unusual exponential shape, whose volume must be
specified as part of the definition of the statistical ensemble.
This gives rise to the Bekenstein-Hawking entropy.

PACS numbers:  0470.Dy, 04.20.Jb, 04.50.Th, 97.60.Lf

\end{abstract}
\pagebreak

The interpretation of the black-hole entropy of Bekenstein and
Hawking, as the logarithm of the number of states in quantum mechanics
or as the logarithm of the available phase space volume in classical
mechanics, has been an ongoing dilemma in gravitation theory. The
purpose of this paper is to propose a solution to this problem, which
may be regarded as formally complete at least
within the limits of validity of Einstein's theory itself.

The central idea is to link the entropy with the surface degrees of
freedom
which give rise to ``off-shell" conical singularities at the horizon
for Euclidean black-hole geometries. This idea was previously
considered by Susskind \cite{susskind}, Ba\~nados, Teitelboim and
Zanelli \cite{BTZ}, and Carlip and Teitelboim \cite{carlip}, but was
not developed to completion. What was missing in the previous
analysis was: (i) A complete action leading upon path integration to the
transition amplitude between an initial and a final state of the
gravitational field in the exterior of a black hole, whose trace
gives the partition function. (ii) A
determination of the entropy in terms of the density of states, both
classical and quantum, i.e., a proper definition of the ensemble at
hand. These two steps are taken here.

The action is obtained by analyzing the boundary terms at the
horizon arising in the variation of the Hamiltonian action for the
gravitational field when the spacetime has the topology of a black
hole. One then finds it necessary to include extra surface fields,
besides the standard ``volume" fields appearing in the action.

Put in classical terms, to account for the black-hole entropy in
terms of the surface fields, it is necessary to determine their
available phase space. It turns out that the requirement of general
covariance fixes the shape of the region of horizon phase space
available to the black hole. The size of the region is governed by a
single dimensionless parameter $\cal N$, which is related to the
range over which a certain horizon canonical coordinate is allowed
to run. This ``volume" $\cal N$ is then held fixed as part of the
definition of the ensemble, {\em in addition} to the mass and
angular momentum (microcanonical ensemble) or {\em in addition}
to the temperature and rotational chemical potential
(grand-canonical ensemble).

One does not see $\cal N$ in the usual calculations because most of
them are done in the semiclassical approximation, where $\cal N$
appears as a pre-factor which is not analyzed, or because the limit
of infinite $\cal N$ is implicitly taken. However, it is crucial for
the present analysis to keep $\cal N$ finite, arbitrary and
fixed.

Turn now to the action principle. Since we will not be concerned with
the interior of the black hole we will regard the Euclidean
formulation as the more fundamental one. This has two important
advantages, namely: (i) The ``one black-hole sector'' is economically
defined by demanding that the complete spacetimes have the topology
$\IR^2 \times S^{D-2}$.  (Extremal black holes require special consideration
\cite{HawHorRoss}, \cite {CT}.  See the comments immediately following
Eq. (\ref{26}) below.)  (ii) The solutions of the classical equations
of motion (the black holes) are regular everywhere and one can
construct an action which has these
solutions as bonafide extrema.

To describe the gravitational field in a convenient manner we
introduce a polar system of coordinates $\rho, \tau$ in $\IR^2$.
For a classical solution (``on-shell") it is convenient to
take the origin $\rho=0$ as the horizon and $\tau$ to be the
Killing time. By abuse
of language, we will call $\rho=0$ the horizon, even away from the
extremum (``off-shell''). The coordinates and metric on $S^{D-2}$
will be denoted by $x^m$ and $\gamma_{mn}$ respectively.

Near the horizon one may take the metric to be given by
\be \label{1}
ds^2 = d\rho^2 + \rho^2\theta^2 d\tau^2 + \gamma_{mn}
(dx^m + N^m d\tau) (dx^n + N^n d\tau) ~~ .
\ee
Here $\theta(x^m,\tau)$ is the proper angle
(proper length divided by proper radius) of an arc in $\IR^2$ of
very small radius, centered at $\rho=0$ and of coordinate angular
opening $d\tau$.

We are interested in keeping the origin fixed, as this is part of
the statement of the problem. For this reason we have set in
(\ref{1}) the lapse and the radial shift at the origin, equal to
zero
\be  \label{2}
N = N^ {\rho} = 0\,, ~~~ {\rm at} ~~ \rho=0 ~~ .
\ee

Note, however, that the projection $N^m(x^p,\tau)$ of the shift
vector on $S^{D-2}(\rho=0)$ is left arbitrary:  we
allow for reparametrizations of the sphere at the origin during the
course of time. This will be of importance in what follows.

Actually to specify an origin one needs more than (\ref{2}), which
hold for any origin.
One may fix a particular origin by demanding that at $\rho=0$ the radial
momentum $\pi^{\rho}_{\rho}$ and the radial derivative
$\gamma^{1/2},_{\rho}$ of the local
area in $S^{D-2}$ vanish,
\be  \label{3}
\pi_{\rho}^{ \rho} = \gamma^{1/2}\,, ~~~ {\rm at} ~~~ {\rho} = 0 ~~ .
\ee
This is nothing but the statement that the differential of
$\gamma^{1/2}$ along $\IR^2$ should vanish,
\be  \label{4}
d_{R^2} \gamma^{1/2} = 0\,, ~~~ {\rm at} ~~ \rho=0 ~~ ,
\ee
that is  $\gamma^{1/2}$ should be stationary.  For a solution of
the equations of motion this happens at the horizon only,
which--incidentally--shows that (\ref {3}) implies (\ref {2}), at
least ``on shell''.

The Euclidean action for the wedge will be taken to be the  sum
\be  \label{5}
I_{\rm wedge}=I_{\rm vol} + I_{\rm hor} ~~ ,
\ee
of the canonical action
\begin{equation} \label{6}
I_{\rm vol} = \int_{\rho >0} d\tau d^{D-1}x \left[\pi^{ij}
\frac{\partial g_{ij}}{\partial d\tau}-N{\cal H} -N^i{\cal H}_i
\right]\,,
\end{equation}
and a surface action for the horizon that we will be determined
presently.

To determine $I_{\rm hor}$ one analyzes the surface terms at
$\rho=0$ in the variation of $I_{\rm vol}$. The surface term
arises from integrations by parts in the variation of $\cal H$
and ${\cal H}_i$, and when $\pi^{\rho}_{\rho}$ and $\gamma^{1/2},_{\rho}$
are held fixed,  is given
by
\be  \label{7}
\int_{\rho =0} d\tau d^{D-1}x \left[-\theta \delta p -N^{m}
\delta \pi_m \right] ~~ ,
\ee
where
\be  \label{8}
p= \left(8\pi G \right)^{1} \gamma^{1/2} \vert _{\rho = 0} ~~ ,
\ee
and
\be   \label{9}
\pi_m= -2 \pi^{\rho}_{m} |_{\rho=0} =2(\gamma/g)^{1/2}
\pi^{i}_{m} n_i |_{\rho=0} ~~ ,
\ee
is the normal-tangential projection at the origin of $\pi^{ij}$ on
$S^{D-2}$, (the vector $n_i$ is the outward normal to
$S^{D-2}$ at a given time $\tau$).

Now, although we might keep $\gamma^{1/2}$ and
$\pi_m$ to be fixed at the origin for all $x,\tau$, it is best to
analyze the action when  their conjugates are left fixed, as this enables
geometrical insight to be gained.  In order to do this we must improve the
action at $\rho=0$ in a manner analogous to the standard
improvement of $I_{\rm vol}$ at large distances which brings in the
mass and angular momentum \cite {regge}.

We thus add to the action a surface term
\be   \label{10}
\int d \tau d^{D-1} x \left (\theta p + N^{m} \pi_{m} \right ) ~~ .
\ee

In the improved action one holds fixed $\theta(x^m,\tau)$ and
$N^m(x^m,\tau)$. Actually, due to the invariance of the action
under surface deformations (changes of the spacetime
coordinates) the action will only depend on the total spatial
diffeomorphism $f^M$ induced by the sequence of shifts
$N^m(x,\tau)$ and on the total arc $\Theta$ measured
perpendicularly to the time constant surfaces. The functions
$f^m$ and $\Theta$ are related to $\theta$ and $N^m$ by
\be   \label{11}
f^m_M \dot{f}^M = N^m ~~ ,
\ee
and
\be   \label{12}
\dot{\Theta} - \Theta,_m f^m_M \dot{f}^M = \theta ~~ .
\ee
Here the dot denotes a partial derivative with respect to $\tau$
and $f^m_M$ is the inverse of the Jacobian matrix $f^M,_m$.

We bring $\Theta$ and $f^M$ into the action by writing $\theta$
and $N^m$ in terms of them through (\ref{11}) and (\ref{12}).
This gives the surface action
\be     \label{13}
I_{\rm hor}=\int d\tau d^{D-2}x \left[p \,\dot{\Theta}+f^m_M
(\pi_m - p \Theta,_m) \dot{f}^{M} \right] ~~ .
\ee
If we denote by $p_M$ the conjugate momentum to $f^M$ we see
from (\ref{12}) that
\be   \label{14}
\pi_m = p \, \Theta,_m + p_M f^{M},_m ~~ ,
\ee
which we recognize as the generator of spatial
reparametrizations of the scalar fields $\Theta, f^{M}$ and
their conjugates $p, \pi_M$. This was to be expected since
(\ref{7}) exhibits $\pi_m$ as conjugate to the surface shift
$N^m$.

It is important to realize that $\pi_m$ is {\em not}
constrained to vanish. This means that the reparametrizations of
the $S^{D-2}$ at $\rho=0$ come in not as gauge symmetries, but rather
as global symmetries (``improper
gauge transformations"). Similarly for the action (\ref{13})
there is no analog of ${\cal H}=0$ at $\rho=0$.


The surface fields $p$ and $\pi_m$ may be regarded as the boundary
data for the constraint equations ${\cal H}_i=0$ and ${\cal H}=0$.
Indeed ${\cal H}_i=0$ considered as a differential equation in the
radial coordinate needs the $\pi_{i}^{\rho}$ to be specified at
$\rho = 0$.  Of these, $\pi^{\rho}_{\rho}$ is set equal to zero on
account of (\ref{3}) leaving $\pi_{m}^{\rho}$ open.  Similarly,
${\cal H}=0$ needs $\gamma^{1/2}$ and
$\gamma^{1/2},_{\rho}$ to be prescribed at the origin. Of these
$\gamma^{1/2},_{\rho}$ is set equal to zero by (\ref{3}) leaving
$\gamma^{1/2}$ open.

The equations of motion derived by extremizing $ I_{\rm vol} +
I_{\rm hor}$ are Einstein's equations and, in addition,
\begin{eqnarray}
\dot{p}- (p N^n),_n=0  \label{15}\,, \\
\dot{\pi}_m-[(\pi_m N^n),_n +\pi_{m,n} N^n +\pi_n N,^n_m] =0\,,  \label{16}
\end{eqnarray}
which express, that up to a spatial reparametrization $p$ and $\pi_m$
are conserved. Here $N^n$ is given by (\ref{11}). Note that the role
of the equations of motion for $\Theta$ and $N^m$ is played by the
definitions (\ref{11}) and (\ref{12}) which -- in the present
formulation -- have been incorporated as identities into the
action. Equations (\ref{11},\ref{12},\ref{15},\ref{16}) reveal the
pairs $(p,\Theta)$, $(\pi_{m},f^M)$ as a sort of action angle variables.

It will be useful in what follows to replace $p$ which is a
scalar density, by an invariant $\hat{p}$ which is strictly
conserved. This is achieved by the canonical transformation
\begin{eqnarray}
\Theta(x,\tau)=\hat{\Theta} \left( f(x,\tau) \right) \equiv
\hat{\Theta} \circ f \label{17} \\
p(x,\tau)= J_f(x,\tau) \hat{p} \left( f(x,\tau) \right) \equiv
J_f \hat{p} \circ f \label{18}
\end{eqnarray}
where $J_f= {\rm det}(f^M_m)$.

Both $\hat{\Theta}$ and $\hat{p}$ are invariant under
reparametrizations. In terms of them the horizon action reads,
\be  \label{19}
I_{\rm hor}=\int d\tau d^{D-2} x \left[ \hat{p}
\,\dot{\hat{\Theta}} + \pi_m f^m_M \dot{f}^{M} \right] ~~ ,
\ee
and Eq. (\ref{12}) becomes just
\be   \label{20}
\dot{\hat{\Theta}} \circ f=\theta ~~ .
\ee

The transition amplitude stems from path integrating the action
obtained by adding (\ref{6}) and (\ref{19}), keeping fixed the fields
$\Theta$, $f$ and $g_{ij}$ at $\tau_1$ and $\tau_2$.  As stated above
the  result will
depend on $\hat{\Theta}, f^M $ only through the combinations
$\hat{\Theta}(\tau_2)-\hat{\Theta}(\tau_1)$ and $ f(\tau_2) \circ
f^{-1}(\tau_1)$. It will be convenient in what follows to pass to the
representation where $\hat{p}(x)$ is diagonal, rather than
$\hat{\Theta}$. This corresponds to adding to the horizon action
(\ref{19}) the total derivative $- \frac{\partial}{\partial
\tau}( \hat{p} \, \hat{\Theta} )$. The corresponding amplitude
will have the form
\be   \label{21}
\delta[\hat{p_2},\hat{p_1}] \, K[ \hat{p_2};f_2 \circ
f^{-1}_1;{\cal G}_2,{\cal G}_1] ~~ ,
\ee
where $\cal G$ denotes the $D-1$-dimensional geometry of
$g_{ij}$ for $\rho>0$.

The amplitude $K$ appearing in (\ref{21}) is the result of path
integrating the volume action $I_{\rm vol}$ given by (\ref{6})
supplemented only by the shift term $N^m\pi_m$, with $\hat{p}(x,\tau)$
taken equal to $\hat{p}(x,\tau_2)$ for all $\tau$. Note that $\pi_m$
is integrated over.

The partition
function is is obtained by taking the trace of (\ref{21}).
This amounts to setting ${\cal G}_2 = {\cal G}_1, f_2=f_1,
\hat{p}_2=\hat{p}_1$ and integrating over their common values.
Observe that setting $f_2=f_1$ may be simply implemented by taking $N^m=0$.
Thus the amplitude $K[ \hat{p_2};1;{\cal G}_2={\cal G}_1]$, where 1
stands for the identity $f (x) = x$, is
obtained by path integrating the
action $I_{\rm vol}$ for the closed
wedge with the local area element of the $S^{D-2}$ at $\rho=0$
held fixed.


In order to preserve the invariance of the theory under
diffeomorphisms of the $S^{D-2}$ at the origin, the integral over
$f_2=f_1$  must be done with an invariant measure over the diffeomorphism
group.  This brings in the (infinite) volume $V({\rm diff}_{S^{D-2}})$ as
an overall factor, making the density of configurations of the field
$\pi_{m}(x)$ proportional to the volume of the diffeomorphism group,
much in the same way as the density of momentum
configurations for a particle in a box is proportional to the volume
of the box.

On the other hand the the formal symbol
$\delta[\hat{p},\hat{p}]$
is the density of states of the field $\hat{p}(x)$ and it is
necessary to give meaning to it. This may be done as follows, by invoking
general covariance.

One wants the partition function -- being a trace -- to be independent
of the choice of basis in the space of states. In the present case, a
choice of basis amounts to a choice of the slicing of spacetime. Now,
if one changes the origin in $\IR^2$ one will change the slicing.
{\em Thus one wants the trace to be independent of the choice of
origin in $\IR^2$}. (Note that this independence from the origin can
only be imposed on the trace and not on the amplitude itself, as the
latter does depend on the initial and final states). But this amounts
to require that the integrand
\be   \label{22}
\delta[\hat{p},\hat{p}]
\exp [\frac{1}{\hbar} I_{\rm vol} \left ({\rm closed~~ wedge} \right)] ~~ ,
\ee
be generally covariant (off shell), which means that one must have
\be   \label{23}
\delta[\hat{p},\hat{p}] = \hat{V} \exp \left({2 \pi\over \hbar}\right)
\int \hat{p}\, d^{D-2}x ~~ ,
\ee
where $\hat{V}$ is a constant.

Indeed the sum
\be   \label{24}
2\pi \int \hat{p}\, d^{D-2}x + I_{\rm vol} = {1 \over 4G}
\int_{p=0} \gamma^{1/2} d^{D-2}x + I_{\rm vol} ~~ ,
\ee
differs from the Hilbert action for the disk by a surface integral at
infinity \cite{BTZ}.

Thus we learn that the number of surface field configurations between
$\pi_m$ and  $\pi_m + d \pi_m, \hat{p}$ and $\hat{p}\ + d \hat{p}$
is given by
\be    \label{25}
{\cal N} \exp \left [\left({2 \pi\over \hbar}\right)
\int \hat{p}\, d^{D-2}x \right] \prod_{x, m}
d\pi_m(x)d\hat p(x) ~~ ,
\ee
with
\be    \label{26}
{\cal N} = V ({\rm diff}_S^{D-2}) \cdot \hat V ~~ ,
\ee
where $\hat V$ is the constant appearing in (\ref{23}).

If matter gauge fields are present additional surface modes appear in
a manner analogous to $\pi_{m}, f^{M}$.  They are of the form
$\pi_{a}, \Lambda^{a}$, where $\pi_{a}$ is the ``electric field
density'' on the horizon and $\Lambda^{a}$ are coordinates on the group
manifold.  The overall $\cal N$ is then  obtained by multiplying the
right side of (\ref {26} ) by $V_G$, the product of the volume of the
internal gauge group with  itself over all points of the sphere.

For extremal black holes the $\IR^2$ part of the metric in (\ref {1}) is
replaced by $e^{2 \rho} d \tau^{2} + d \rho^{2}$ and the origin, at
$\rho = -\infty$, is not on the manifold.  As a result $(\gamma^{1/2},
\Theta)$ disappears as an independent canonical pair of surface
variables ($\Theta$ may be thought of as being zero and $\gamma^{1/2}$
becomes a function of $\pi_m$ and the gauge momenta $\pi_a$).
Therefore the factor $\delta [\hat p, \hat p]$ is absent from the
trace and $\cal N$ is given just by $V ({\rm diff}_S^{D-2}) \cdot V_G$.

In order to understand the meaning of (\ref {23}) and -- in particular
-- that of the constant $\hat V$ in (\ref {26} ) it helps to go back
to the horizon action (\ref {19}) and calculate the volume element
$\omega$ in the phase space of ($\hat p, \hat \Theta$), ($\pi_m,
f^M$).  One finds

\be    \label{27}
\omega = \prod_{x,m,M} \left [d \hat {p} (x) \wedge d {\hat \Theta} (x) \wedge
d {\pi_m} (x) \wedge df^{M} (x)\right] J_{f}^{-1} (x) ~~ .
\ee
Now, it is immediate to see that

\be    \label{28}
Df = \prod_{x,v} df^{1} (x) \wedge ... \wedge df^{D-2} (x) J_{f}^{-1} (x) ~~,
\ee
is an invariant measure over the diffeomorphism group (one has $D (foh) =
Df$ for all $h$).  Thus the factor $V({\rm diff}_{S^{D-2}})$ comes in
straight away.

There remains to analyze the integral over $\hat p$ and $\hat \Theta$.
The question to be asked is, how can one achieve that the phase space
volume between $\hat {p} (x) $ and $\hat {p} (x) + d {\hat p} (x)$ be of
the form (\ref {23} )?  The answer is simple: one needs to properly
adjust the region of integration for $\hat {\Theta} (x)$.

To see this it is useful to imagine expanding $\hat {\Theta}$ and
$\hat p$ in terms of a set of orthonormal functions
over the sphere (``spherical-harmonics'')labeled by a collective
index $i$.  Denote by
$\hat {\Theta}_0, \hat {p}_0$ the ``zero modes'', that is, the
integrals of $\hat \Theta$ and $\hat p$ over the sphere.  For
fixed $\hat {p}_i$ ($i \geq 0$), the region of integration for $\hat
{\Theta}_i$ is defined as follows:  the zero mode is confined to an
``exponential box'':
\be    \label{29}
0 \leq \hat{\Theta}_{0} \leq \hat{L}_{0} \exp \left ({2 \pi \over \hbar} {\hat
p}_0 \right ) ~~ ,
\ee
and the higher modes are contained in an ordinary rectangular box
\be    \label{30}
0 \leq \hat{\Theta}_{i} \leq \hat{L}_{i}, ~~~~~ i \geq 1 ~~ .
\ee

If we calculate the volume in $\hat {\Theta}$, $\hat
p$ space between $\hat {p}_i$ and $d \hat {p}_i$ we find using(\ref
{29}) and (\ref {30}) a
density of the form (\ref {23}), with
\be    \label{31}
\hat {V} = \prod_{i \geq 0} \hat L_i ~~ .
\ee

Thus we see that the exponentially enhanced density (\ref {23}) may be
thought of as due to a peculiar sort of ``box'' for the zero mode of
the $\hat {p}, \hat {\Theta}$ degrees of freedom.  The shape of the
box is mandated by general covariance.  Therefore, we might call (\ref {23})
the density in the ``generally covariant ensemble.''

Note that although the preceding analysis is purely classical in
language, Planck's constant $\hbar$ is built in the definition (\ref {29} ) of
the ensemble which, therefore, is inherently quantum mechanical.

One may also obtain the density (\ref {23}) by counting eigenvalues.
To this effect perform first the canonical transformation
\be    \label{32}
P_{0} = {\hbar \over 2 \pi} \exp {2 \pi \over \hbar} \hat {p}_0, Q_{0}
= \hat {\Theta}_{0} \exp \left (-{2 \pi \over \hbar} \hat p_{0}\right ) ~~ ,
\ee
which involves $\hbar$.  Then (\ref {29}) says that the coordinate
$Q_0$ is confined to an ordinary box of length $\hat L_0$, $ 0 \leq Q_{0}
\leq \hat {L_0}$.  If one imposes periodic boundary conditions on this box,
the eigenvalues $(P_0)_n$ of $P_0$ are equally spaced: $(P_0)_n =
\left (2 \pi \hbar/ \hat L_0 \right)n$, with  $n$ integer, which means
that the  eigenvalues $(\hat p_0)_n$ of $\hat p_0$ are logarithmically
spaced,

\be    \label{33}
\left(\hat p_0 \right)_n = {\hbar \over 2 \pi} \log  {4 \pi^2 \over
\hat L_0}n ~~ ,
\ee
yielding for large $\hat L_0$ a continuous spectral density of the
exponential form (\ref {23}).

Lastly, observe that for a black hole classical solution $\hat p$ and
$\pi_m$ are expressed in terms of $M$ and $J$ and $I_{\rm vol}$
vanishes, therefore to the lowest orders in $\hbar$ the entropy reads
\be    \label{34}
S [M, J, {\cal N} ] = {1 \over 4G \hbar } {A_+} \left (J,M \right) + \log \cal
{N} ~~ ,
\ee
where $A_+$ is the area of the horizon.

As long as $\cal N$ is kept fixed, the log $\cal N$ contribution to
the entropy has no implications for thermodynamics.  However,
according to (\ref {26}) one has,
\be    \label{35}
\log {\cal {N}} = \log V \left ({\rm diff}_{S^{D-2}} \right)+ \log \hat V ~~ ,
\ee
and it would seem reasonable that -- even after a cutoff is introduced to
regularize it -- $V ({\rm diff}_{S^{D-2}})$ should be taken to be a
universal constant of topological origin.  The same applies to the
additional $\log V_G$ appearing in (\ref {35}) when gauge matter
fields are brought in.

It does not seem equally evident though, that $\hat V$ should be
treated as universally fixed.  Indeed one might imagine, for example,
that black holes formed in different ways could acquire $\hat V's$
differing from each other by a finite amount.  If this were the case, a new
thermodynamic parameter for black holes would appear.  At the moment
of this writing this remains just a speculative note.

The author is grateful to M. Ba\~{n}ados, S. Carlip, D. Gross, F.
Wilczek, and J. Zanelli for helpful discussions.  This work was
supported in part by Grant No.  1940203 of FONDECYT (Chile).
Institutional support to CECS from a group of Chilean private
companies (COPEC, CGE, Empresas CMPC, ENERSIS, MINERA LA ESCONDIDA,
IBM and XEROX) is also acknowledged.
\pagebreak

\end{document}